\documentclass[prd,showpacs,showkeys,amsmath,amssymb]{revtex4}
\usepackage{hyperref,slashed,graphicx,color}
\begin{document}
\title{A Note on analytic formulas of Feynman propagators in position space}
\author{Hong-Hao Zhang$^1$}
\author{Kai-Xi Feng$^1$}
\author{Si-Wei Qiu$^1$}
\author{An Zhao$^1$}
\author{Xue-Song Li$^2$}

\affiliation{$^1$School of Physics and Engineering, Sun Yat-Sen
University, Guangzhou 510275, China\\
$^2$Science College, Hunan Agricultural University, Changsha 410128,
China}

\begin{abstract}
In this paper, we correct an inaccurate result of previous works on
the Feynman propagator in position space of a free Dirac field in
$(3+1)$-dimensional spacetime, and we derive the generalized
analytic formulas of both the scalar Feynman propagator and the
spinor Feynman propagator in position space in arbitrary
$(D+1)$-dimensional spacetime, and we further find a recurrence
relation among the spinor Feynman propagator in $(D+1)$-dimensional
spacetime and the scalar Feynman propagators in $(D+1)$-, $(D-1)$-
and $(D+3)$-dimensional spacetimes.
\end{abstract}

\maketitle

\section{Introduction}
Although one cannot adopt the extreme view that the set of all
Feynman rules represents the full theory of quantized fields, the
approach of the Feynman graphs and rules plays an important role in
perturbative quantum field theories. For a generic quantum theory
involving interacting fields, the set of its Feynman rules includes
some vertices and propagators. While for a free field theory, there
is only one graph, that is, the Feynman propagator, in the set of
its Feynman rules. Thus the Feynman propagator describes most, if
not all, of the physical content of the free field theory. It is
true that the real world is not governed by any free field theory;
however, this kind of theory is the base of a perturbatively
interacting field theory, and the issues of a free theory are
usually in the simplest situation and thus will be attempted to
study at the first step of investigations.

It has been mentioned in
Ref.\cite{Greiner:1992bv,DeWitt:1965jb,Peskin:1995ev,
Huang:1998qk,Zee:2003mt,Bekenstein:1981xe,McKay:1996th,Antonsen:1996kc}
that in $(3+1)$-dimensional spacetime, after integrating over
momentum, the Feynman propagator of a free Klein-Gordon field can be
expressed in terms of Bessel or modified Bessel functions, which
depends on whether the separation of two spacetime points is
timelike or spacelike. By changing variables to hyperbolic functions
and using the integral representation of the Hankel function of
second kind, the authors of Ref.\cite{Greiner:1992bv} derived the
full analytic formulas of the Feynman propagors of free Klein-Gordon
and Dirac fields in $(3+1)$-dimensional spacetime. And the
expressions of the Feynman propagators of a free Klein-Gordon field
in $(1+1)$- and that in $(2+1)$-dimensional spacetime can be found
in Ref.\cite{DiSessa:1974xd} and \cite{Gutzwiller2003},
respectively.  However, in Ref.\cite{Greiner:1992bv} the expression
for the Feynman propagator of a Dirac spinor field is inaccurate,
since there is at least a redundant term in their results. In this
paper we will show that the term actually vanishes and we will give
the correct expression. Furthermore, we will generalize the results
of previous works and derive the full analytic formulas of the
Feynman propagators in position space of, respectively, the
Klein-Gordon scalar and the Dirac spinor in arbitrary
$(D+1)$-dimensional spacetime. Eventually we will find an
interesting recurrence relation between the spinor Feynman
propagator in $(D+1)$-dimensional spacetime and the scalar Feynman
propagators in $(D+1)$- and alternate-successive dimensional
spacetimes.

This paper is organized as follows. In Section
\ref{section-klein-gordon}, after briefly reviewing the derivation
of the analytic formulas of the Feynman propagators of a free
Klein-Gordon field in $(1+1)$- and $(2+1)$-dimensional spacetime, we
will compute, once and for all, the scalar Feynman propagator in
$(D+1)$-dimensional spacetime, and we will compare our results for
the case $D=1,2,3$ with the results of previous works. Although the
method we will use is different from that used in
Ref.\cite{Greiner:1992bv}, we will see that both the results are
consistent with each other. In Section \ref{section-dirac} we will
make use of the obtained formula of the scalar Feynman propagator to
compute the expression of the spinor Feynman propagator in
$(D+1)$-dimensional spacetime, and we will also compare this result
for $D=3$ with that of Ref.\cite{Greiner:1992bv}, and we will show
that one additional term in Ref.\cite{Greiner:1992bv} actually has
no contribution and the method they used to prove the term nonzero
was inappropriate. And we will further derive a recurrence relation
between the Feynman propagators in position space of the Dirac
spinor field and the Klein-Gordon scalar fields in three
alternate-successive dimensional spacetimes. The last section is
devoted to conclusions.

\section{Feynman propagator of Klein-Gordon Theory \label{section-klein-gordon}}

Following the notation of Ref.\cite{Peskin:1995ev,Zee:2003mt}, we
write the Feynman propagator of a free Klein-Gordon field
$\phi(x)\equiv\phi(t,\vec{x})$ in $(D+1)$-dimensional spacetime as
the time-ordered two-point correlation function:
\begin{eqnarray}
D_F(x)\equiv\langle0|T\phi(x)\phi(0)|0\rangle=\theta(x^0)D(x)+\theta(-x^0)D(-x)\;,
\end{eqnarray}
with the unordered two-point correlation function
\begin{eqnarray}
D(x)\equiv\langle0|\phi(x)\phi(0)|0\rangle
=\int\frac{d^D\vec{p}}{(2\pi)^D}\frac{1}{2E_{\vec{p}}}
e^{-i(E_{\vec{p}}t-\vec{p}\cdot\vec{x})}\;.
\end{eqnarray}
Combining the above two equations gives
\begin{eqnarray}
D_F(t,\vec{x})=D(|t|,\vec{x})=\int\frac{d^D\vec{p}}{(2\pi)^D}\frac{1}{2E_{\vec{p}}}
e^{-i(E_{\vec{p}}|t|-\vec{p}\cdot\vec{x})}\;.\label{eq-Feyn-Propa-1}
\end{eqnarray}
Thus, we can obtain the analytic formula of the Feynman propagator
in position space by integrating over the $D$-dimensional momentum
in the expression of $D(|t|,\vec{x})$. This integral in
$D$-dimensional Euclidean space can be evaluated by changing the
variables from Cartesian coordinates to spherical coordinates. Since
the angular integral parts look a little different in $(1+1)$-,
$(2+1)$- and general $(D+1)$-dimensional (for $D\geq3$) spacetimes,
in order to be more careful in our derivation, let us consider these
situations case by case. Eventually we will show that the general
result of $(D+1)$-dimensional spacetime holds for $D=1,2$ as well.

\subsection{$(1+1)$-dimensional spacetime}

When the spatial dimension $D=1$, eq.\eqref{eq-Feyn-Propa-1} becomes
\begin{eqnarray}
D_F(t,r)=\int_{-\infty}^{+\infty}\frac{dp}{2\pi}\frac{1}{2E}e^{-i(E|t|-pr)}\;,\qquad
\mbox{with}\quad E=\sqrt{p^2+m^2}
\end{eqnarray}
Using the substitution $E=m\cosh\eta$, $p=m\sinh\eta$ (with
$-\infty<\eta<\infty$) in the above integral, we have
\begin{eqnarray}
D_F(t,r)=\frac{1}{4\pi}\int_{-\infty}^{+\infty}d\eta
e^{-im(|t|\cosh\eta-r\sinh\eta)}
\end{eqnarray}
Due to the Lorentz invariance, the Feynman propagator can depond
only on the interval $x^2\equiv t^2-r^2$. If the interval is
timelike, $x^2>0$, we can make a Lorentz transformation such that
$x$ is purely in the time-direction:
$x^0=\theta(t)\sqrt{t^2-r^2-i\epsilon}$, $r=0$. Note that $\sqrt{s}$
is not a single-valued-function of $s$ and here and henceforth the
cut line in the complex plane of $s$ is chosen to be the negative
real axis, and the negative infinitesimal imaginary part,
$-i\epsilon$, is because of the Feynman description of the Wick
rotation, i.e., $x^2\to x^2-i\epsilon$ in position space and
correspondingly $k^2\to k^2+i\epsilon$ in momentum space. Thus,
\begin{eqnarray}
D_F(t,r)&=&\frac{1}{4\pi}\int_{-\infty}^{+\infty}d\eta
e^{-im\sqrt{t^2-r^2-i\epsilon}\cosh\eta}\nonumber\\
&=&-\frac{i}{4}
H_0^{(2)}(m\sqrt{t^2-r^2-i\epsilon})\label{eq-1p1-timelike}
\end{eqnarray}
where $H_0^{(2)}(x)$ is the Hankel function of second kind, and
where we have used the identity in \# 3.337 of
Ref.\cite{tablesofintegrals},
\begin{eqnarray}
\int_{-\infty}^{+\infty}d\eta e^{-i\beta\cosh\eta}=-i\pi
H_0^{(2)}(\beta)\;,\qquad (-\pi<\arg\beta<0)\label{math-identity-1}
\end{eqnarray}
Likewise, if the interval is spacelike, $x^2<0$, we have
\begin{eqnarray}
D_F(t,r)&=&\frac{1}{4\pi}\int_{-\infty}^{+\infty}d\eta
e^{im\sqrt{r^2-t^2+i\epsilon}\sinh\eta}\nonumber\\
&=&\frac{1}{2\pi}K_0(m\sqrt{r^2-t^2+i\epsilon})\label{eq-1p1-spacelike}
\end{eqnarray}
where $K_0(x)$ is the modified Bessel function, and where we have
used the identity in \# 3.714 of Ref.\cite{tablesofintegrals},
\begin{eqnarray}
\int_0^\infty d\eta \cos(\beta\sinh\eta)=K_0(\beta)\;,\qquad
(-\frac{\pi}{2}<\arg \beta<\frac{\pi}{2})\label{math-identity-2}
\end{eqnarray}
It is worthwhile noting that the $-i\epsilon$ description assures
the proper phase angles of $\sqrt{x^2-i\epsilon}$ in
eq.\eqref{eq-1p1-timelike} and $\sqrt{-x^2+i\epsilon}$ in
eq.\eqref{eq-1p1-spacelike}, respectively, so that the mathematical
identities \eqref{math-identity-1} and \eqref{math-identity-2} can
happen to be applied to figure out these two expressions. In the
limit of $x^2\to 0$, both eqs. \eqref{eq-1p1-timelike} and
\eqref{eq-1p1-spacelike} are divergent and are approaching
\begin{eqnarray}
&&\lim_{x^2\to 0}-\frac{i}{4}
H_0^{(2)}(m\sqrt{x^2-i\epsilon})\sim\frac{1}{4\pi}\ln\frac{1}{x^2-i\epsilon}\\
&&\lim_{x^2\to 0}\frac{1}{2\pi}K_0(m\sqrt{-x^2+i\epsilon})\sim
\frac{1}{4\pi}\ln\frac{1}{x^2-i\epsilon}\label{eq-1p1-spacelike-1}
\end{eqnarray}
which are of the same form and do not depend on the mass $m$, and
which can be recognized as the scalar Feynman propagator on the
lightcone. The above expression is indeed the Feynman propagator of
a massless scalar field,
\begin{eqnarray}
D_F(x)=\frac{1}{4\pi}\ln\frac{1}{x^2-i\epsilon}\;,\qquad
\mbox{for}\quad m=0 \label{eq-1p1-massless}
\end{eqnarray}
In summary, the scalar Feynman propagator in position space may be
written in a compact way as
\begin{eqnarray}
D_F(x)=\theta(x^2)\bigg(-\frac{i}{4}
H_0^{(2)}(m\sqrt{x^2-i\epsilon})\bigg)
+\theta(-x^2)\frac{1}{2\pi}K_0(m\sqrt{-x^2+i\epsilon})\label{eq-1p1-summary}
\end{eqnarray}
where the theta function $\theta(x)$ is defined as
\begin{eqnarray}
\theta(x)=\int_{-\infty}^x \delta(y)d{y}= \left\{
\begin{array}{ll}
1\;,\quad & (x>0)\\
0\;,\quad & (x<0)
\end{array}
\right.
\end{eqnarray}
and the value of $\theta(x=0)$ depends on whether the argument $x$
is approaching to $0$ from the positive or negative real axis, that
is, $\theta(0^+)=1$ and $\theta(0^-)=0$.

\subsection{$(2+1)$-dimensional spacetime}

In $(2+1)$-dimensional spacetime, the Feynman propagator of a free
scalar field is
\begin{eqnarray}
D_F(t,r)=\frac{1}{(2\pi)^2}\int_0^{2\pi}d\theta\int_0^\infty dp
\frac{p}{2E}e^{-i(E|t|-pr\cos\theta)}\;,\qquad \mbox{with}\quad
E=\sqrt{p^2+m^2}
\end{eqnarray}
Using the similar calculation procedure, we can obtain the analytic
formula of the scalar Feynman propagator in $(2+1)$-dimensional
spacetime as follows
\begin{eqnarray}
D_F(x)=\theta(x^2)\frac{-i}{4\pi
\sqrt{x^2-i\epsilon}}e^{-im\sqrt{x^2-i\epsilon}}+\theta(-x^2)\frac{1}{4\pi
\sqrt{-x^2+i\epsilon}}e^{-m\sqrt{-x^2+i\epsilon}}\label{eq-2p1-summary}
\end{eqnarray}
Eqs. \eqref{eq-1p1-summary} and \eqref{eq-2p1-summary} agree well
with the results of previous works
\cite{DiSessa:1974xd,Gutzwiller2003}.

\subsection{$(D+1)$-dimensional spacetime (for $D\geq 3$)}

\begin{figure}[t!]
\begin{center}
\includegraphics[height=8cm]{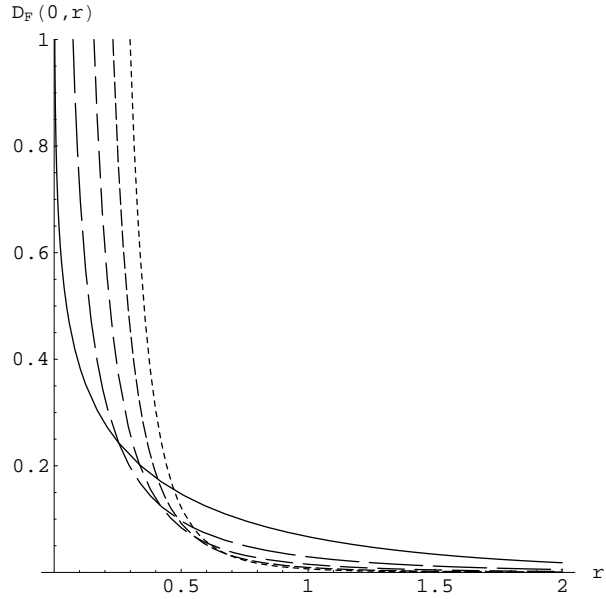}
\end{center}
\caption[1]{\label{fig-spacelike-DF}The scalar Feynman propagator
$D_F(0,r)$ with spacelike separation $r$ in $(D+1)$-dimensional
spacetime, where we have set the mass parameter $m=1$; and the
solid-line corresponds to $D=1$, while the dashed-lines, from long
to short, correspond to $D=2,3,4,5$, respectively.}
\end{figure}

\begin{figure}[t!]
\begin{center}
\includegraphics[height=8cm]{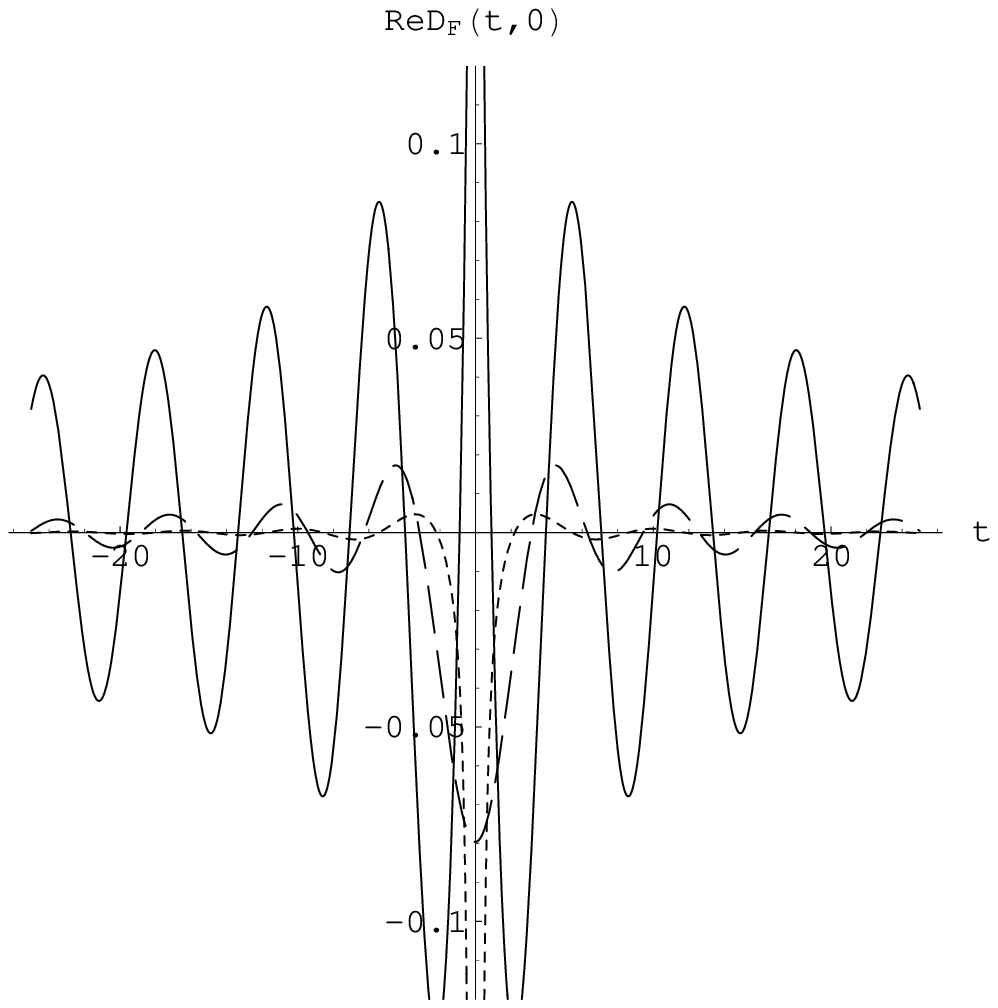}
\includegraphics[height=8cm]{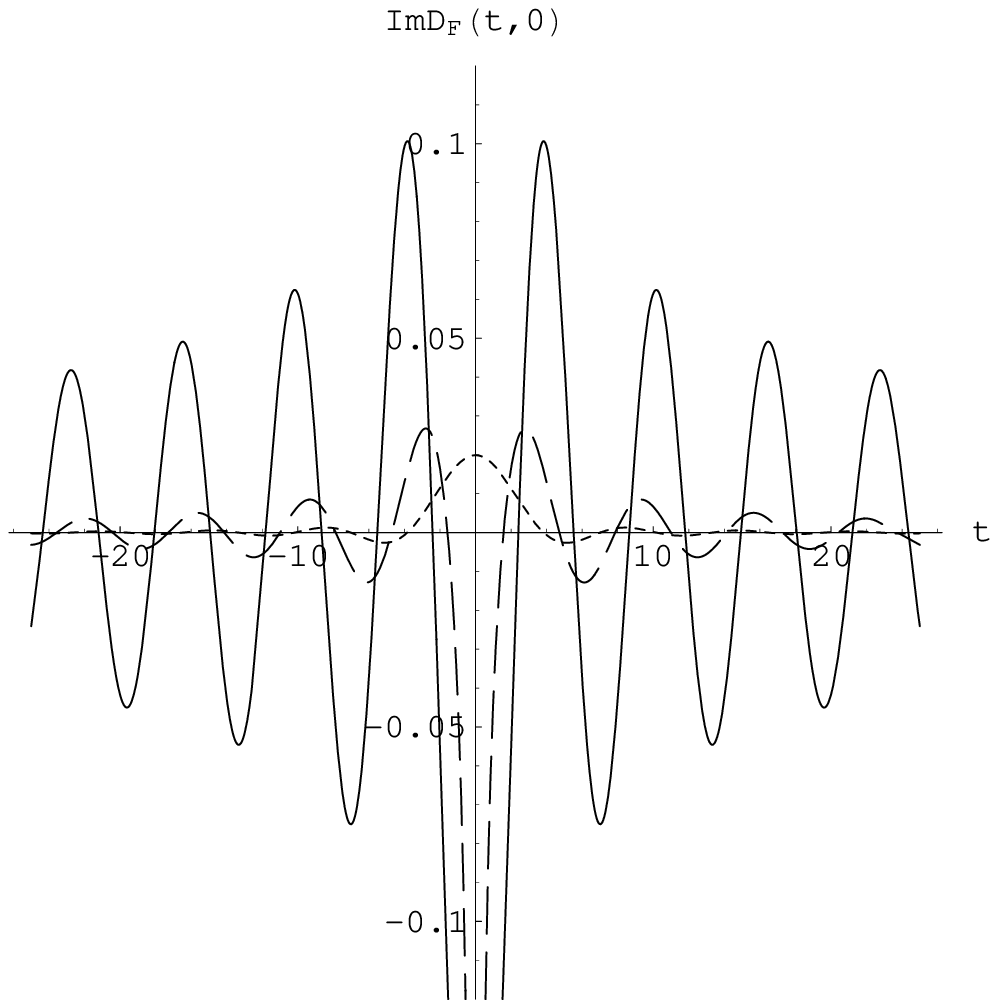}
\end{center}
\caption[1]{\label{fig-timelike-DF}The real and imaginary parts of
the scalar Feynman propagator $D_F(t,0)$ with timelike separation
$t$ in $(D+1)$-dimensional spacetime, where we have set the mass
parameter $m=1$; and the solid-line corresponds to $D=1$, the long
dashed-line $D=2$, and the short dashed-line $D=3$.}
\end{figure}

Now, let us proceed to compute the scalar Feynman propagator in
$(D+1)$-dimensional spacetime (for $D\geq 3$). Since the method we
will use in the following differs from that used in
Ref.\cite{Greiner:1992bv}, let us wait to see whether the results
from these two approaches are consistent or not. Changing the
variables from Cartesian coordinates to spherical coordinates,
eq.\eqref{eq-Feyn-Propa-1} becomes
\begin{eqnarray}
D_F(t,r)=\frac{1}{(2\pi)^D}\frac{2\pi^{\frac{D-1}{2}}}{\Gamma(\frac{D-1}{2})}
\int_0^{\pi}\sin^{D-2}\theta d\theta \int_0^{\infty}dp
\frac{p^{D-1}}{2E} e^{-i(E|t|-pr\cos\theta )}\;,\qquad
\mbox{with}\quad E=\sqrt{p^2+m^2}\label{eq-Feyn-Propa-2}
\end{eqnarray}
To evaluate the above integral, we need to figure out the angular
integral $\int_0^{\pi}d\theta\sin^{D-2}\theta e^{ipr\cos\theta}$.
From the formula \# 3.387 of Ref.\cite{tablesofintegrals},
\begin{eqnarray}
\int_{-1}^1dx(1-x^2)^{\nu-1}e^{i\mu
x}=\sqrt{\pi}\bigg(\frac{2}{\mu}\bigg)^{\nu-\frac{1}{2}}\Gamma(\nu)J_{\nu-\frac{1}{2}}(\mu)\;,
\quad (~\mathrm{Re}~\nu >0~)
\end{eqnarray}
we can easily find that
\begin{eqnarray}
\int_0^\pi d\theta \sin^k\theta e^{ipr\cos\theta}=
\sqrt{\pi}\bigg(\frac{2}{pr}\bigg)^{\frac{k}{2}}
\Gamma(\frac{k+1}{2})J_{\frac{k}{2}}(pr)\label{integral-formula-2}
\end{eqnarray}
Then, substituting eq.\eqref{integral-formula-2} with $k=D-2$ into
eq.\eqref{eq-Feyn-Propa-2}, we obtain
\begin{eqnarray}
D_F(t,r)=\frac{1}{2(2\pi)^{\frac{D}{2}}r^{\frac{D}{2}-1}}
\int_0^{\infty}dp
\frac{p^{\frac{D}{2}}}{E}J_{\frac{D}{2}-1}(pr)e^{-iE|t|}\;,\qquad
\mbox{with}\quad E=\sqrt{p^2+m^2}
\end{eqnarray}
which, by changing variable of integration to $x=E/m$, leads to
\begin{eqnarray}
D_F(t,r)=\frac{m^{\frac{D}{2}}}{2(2\pi)^{\frac{D}{2}}r^{\frac{D}{2}-1}}
\int_1^\infty dx (x^2-1)^{\frac{1}{2}(\frac{D}{2}-1)}
J_{\frac{D}{2}-1}(mr\sqrt{x^2-1}) e^{-im|t|x}\label{eq-Feyn-Propa-3}
\end{eqnarray}
To compute the above integral, we can make the analytical
continuation of the 2nd formula of \# 6.645 of
Ref.\cite{tablesofintegrals},
\begin{eqnarray}
\int_1^\infty dx(x^2-1)^{\frac{1}{2}\nu}e^{-\alpha x}J_\nu
(\beta\sqrt{x^2-1})=\sqrt{\frac{2}{\pi}}\beta^\nu
(\alpha^2+\beta^2+i\epsilon)^{-\frac{1}{2}\nu-\frac{1}{4}}
K_{\nu+\frac{1}{2}}(\sqrt{\alpha^2+\beta^2+i\epsilon})
\end{eqnarray}
and obtain the following identity:
\begin{eqnarray}
\int_1^\infty dx(x^2-1)^{\frac{1}{2}\nu}e^{-ia x}J_\nu
(b\sqrt{x^2-1})=\left\{
\begin{array}{ll}
\sqrt{\frac{2}{\pi}}b^\nu
(b^2-a^2+i\epsilon)^{-\frac{1}{2}\nu-\frac{1}{4}}
K_{\nu+\frac{1}{2}}(\sqrt{b^2-a^2+i\epsilon})\;, &\quad
(b>a>0)\\
\sqrt{\frac{\pi}{2}}b^\nu
\frac{(-i)^{2(\nu+1)}}{(\sqrt{a^2-b^2-i\epsilon})^{\nu+\frac{1}{2}}}
H_{\nu+\frac{1}{2}}^{(2)}(\sqrt{a^2-b^2-i\epsilon})\;, & \quad
(a>b>0)
\end{array}
\right.\label{integral-formula-3}
\end{eqnarray}
Substituting eq.\eqref{integral-formula-3} (with
$\nu=\frac{D}{2}-1\,,~a=mt\,,~b=mr$) into
eq.\eqref{eq-Feyn-Propa-3}, we obtain
\begin{eqnarray}
&&D_F(t,r)=\frac{(-i)^Dm^{\frac{D-1}{2}}}{2^{\frac{D+3}{2}}\pi^{\frac{D-1}{2}}
(t^2-r^2-i\epsilon)^{\frac{D-1}{4}}}H_{\frac{D-1}{2}}^{(2)}(m\sqrt{t^2-r^2-i\epsilon})\;,\qquad
(\mbox{for}\quad t^2-r^2>0)\\
&&D_F(t,r)=\frac{m^{\frac{D-1}{2}}}{(2\pi)^{\frac{D+1}{2}}(r^2-t^2+i\epsilon)^{\frac{D-1}{4}}}
K_{\frac{D-1}{2}}(m\sqrt{r^2-t^2+i\epsilon})\;,\qquad
(\mbox{for}\quad r^2-t^2>0)
\end{eqnarray}
In the limit of $x^2\to 0$, the above two formulas are approaching
to a common asymptotic expression, that is, the scalar Feynman
propagator on the lightcone:
\begin{eqnarray}
D_F(x)\sim\frac{\Gamma(\frac{D-1}{2})}{4\pi^{\frac{D+1}{2}}}
\bigg(-\frac{1}{x^2-i\epsilon}\bigg)^{\frac{D-1}{2}}\;,\qquad
(\mbox{for}\quad x^2\to 0)
\end{eqnarray}
which is indeed the exact formula of the Feynman propagator of a
massless scalar field. In summary, the full analytic expression of
the scalar Feynman propagator in $(D+1)$-dimensional spacetime is
given by
\begin{eqnarray}
D_F(x)&=&\theta(x^2)\frac{(-i)^Dm^{\frac{D-1}{2}}}{2^{\frac{D+3}{2}}\pi^{\frac{D-1}{2}}
(x^2-i\epsilon)^{\frac{D-1}{4}}}H_{\frac{D-1}{2}}^{(2)}(m\sqrt{x^2-i\epsilon})\nonumber\\
&&+\theta(-x^2)\frac{m^{\frac{D-1}{2}}}{(2\pi)^{\frac{D+1}{2}}(-x^2+i\epsilon)^{\frac{D-1}{4}}}
K_{\frac{D-1}{2}}(m\sqrt{-x^2+i\epsilon})\label{eq-Dp1-summary}
\end{eqnarray}
In particular, taking $D=3$, it follows from the above equation that
\begin{eqnarray}
D_F(x)=\theta(x^2)\frac{im}{8\pi\sqrt{x^2-i\epsilon}}H_1^{(2)}(m\sqrt{x^2-i\epsilon})
+\theta(-x^2)\frac{m}{4\pi^2\sqrt{-x^2+i\epsilon}}K_1(m\sqrt{-x^2+i\epsilon})
\end{eqnarray}
which is consistent with the results in $(3+1)$-dimensional
spacetime of Ref.\cite{Greiner:1992bv}. Moreover,
eq.\eqref{eq-Dp1-summary} holds not only for $D\geq3$, but also for
$D=1,2$. Noting the facts that
\begin{eqnarray}
H_{\frac{1}{2}}^{(2)}(x)=i\sqrt{\frac{2}{\pi x}}e^{-ix}\;,\qquad
K_{\frac{1}{2}}(x)=\sqrt{\frac{\pi}{2x}}e^{-x}\;,
\end{eqnarray}
it can be easily verified that if the spatial dimension is taken to
be $D=1,2$, eq.\eqref{eq-Dp1-summary} will reduce to
eqs.\eqref{eq-1p1-summary},\eqref{eq-2p1-summary}, respectively.
Therefore, in the following we will use eq.\eqref{eq-Dp1-summary} to
describe the scalar Feynman propagator in $(D+1)$-dimensional
spacetime for $D\geq 1$. The shapes of the scalar Feynman
propagators with spacelike or timelike separations in different
dimensional spacetime are shown in Figs.\ref{fig-spacelike-DF} and
\ref{fig-timelike-DF}, respectively. The figures exhibit that in any
dimensional spacetime, the spacelike propagation amplitude is
dominated by the exponential decay, while the timelike propagation
amplitude behaves as the damped oscillation; and in both cases the
more dimension of spacetime, the more rapidly the propagation
amplitude decreases.


\section{Feynman propagator of Dirac Theory \label{section-dirac}}

In this section, let us calculate the analytic formula of the
Feynman propagator in position space of a free Dirac spinor field in
$(D+1)$-dimensional spacetime. Since we have obtained the exact
expression of the scalar Feynman propagator in any dimensional
spacetime, eq.\eqref{eq-Dp1-summary}, it is straightforward to get
the expression of the spinor Feynman propagator by means of the
relation $S_F(x)=(i\slashed{\partial}+m)D_F(x)$. The result we
obtain is
\begin{eqnarray}
S_F(x)&=&\theta(x^2)\frac{(-i)^{D-1}m^{\frac{D+1}{2}}\slashed{x}}
{2^{\frac{D+5}{2}}\pi^{\frac{D-1}{2}}
(x^2-i\epsilon)^{\frac{D+1}{4}}}
\bigg[H_{\frac{D-3}{2}}^{(2)}(m\sqrt{x^2-i\epsilon})
-H_{\frac{D+1}{2}}^{(2)}(m\sqrt{x^2-i\epsilon})\bigg]\nonumber\\
&&+\theta(-x^2)\frac{im^{\frac{D+1}{2}}\slashed{x}}
{2^{\frac{D+3}{2}}\pi^{\frac{D+1}{2}}(-x^2+i\epsilon)^{\frac{D+1}{4}}}
\bigg[K_{\frac{D-3}{2}}(m\sqrt{-x^2+i\epsilon})
+K_{\frac{D+1}{2}}(m\sqrt{-x^2+i\epsilon})\bigg]\nonumber\\
&&-\frac{(D-1)}{2}\frac{i\slashed{x}}{x^2-i\epsilon}D_F(x)+mD_F(x)\label{eq-Dp1-Dirac}
\end{eqnarray}
where $\slashed{x}\equiv \gamma^\mu x_\mu$, and where we have used
the following recurrence relations of the Hankel function
$H_\nu^{(2)}(x)$ and the modified Bessel function $K_\nu(x)$:
\begin{eqnarray}
&&\frac{d}{dx}H_\nu^{(2)}(x)=\frac{1}{2}\big[H_{\nu-1}^{(2)}(x)-H_{\nu+1}^{(2)}(x)\big]\\
&&\frac{d}{dx}K_\nu(x)=-\frac{1}{2}\big[K_{\nu-1}(x)+K_{\nu+1}(x)\big]
\end{eqnarray}
In particular, when the spatial dimension $D=3$,
eq.\eqref{eq-Dp1-Dirac} becomes
\begin{eqnarray}
S_F(x)&=&-\theta(x^2)\frac{m^2\slashed{x}}{16\pi(x^2-i\epsilon)}
\bigg[H_{0}^{(2)}(m\sqrt{x^2-i\epsilon})
-H_{2}^{(2)}(m\sqrt{x^2-i\epsilon})\bigg]\nonumber\\
&&+\theta(-x^2)\frac{im^2\slashed{x}}{8\pi^2(-x^2+i\epsilon)}
\bigg[K_{0}(m\sqrt{-x^2+i\epsilon})
+K_{2}(m\sqrt{-x^2+i\epsilon})\bigg]\nonumber\\
&&-\frac{i\slashed{x}}{x^2-i\epsilon}D_F(x)+mD_F(x)\label{eq-3p1-Dirac}
\end{eqnarray}
which is a little different from the results of
Ref.\cite{Greiner:1992bv}, besides the less important factor of $i$
owing to the convention that $D_F(x)$ here equals to $i\Delta_F(x)$
there. The essential difference between our results and those of
Ref.\cite{Greiner:1992bv} lies in that there is an additional term
multiplied by $\delta(x^2)$ in that book, which is proportional to
eq.(34) in page 80 of Ref.\cite{Greiner:1992bv}. However, we find
that this term is redundant, since its proportional factor can be
shown to vanish as follows:
\begin{eqnarray}
&&\hspace{-0.5cm}\lim_{x^2\to
0}\bigg[\frac{1}{\sqrt{x^2-i\epsilon}}H_1^{(2)}(m\sqrt{x^2-i\epsilon})
-\frac{i}{\sqrt{-x^2+i\epsilon}}H_1^{(2)}(-im\sqrt{-x^2+i\epsilon})\bigg]\nonumber\\
&\sim&\frac{1}{\sqrt{x^2-i\epsilon}}\frac{i}{\pi}\frac{2}{m\sqrt{x^2-i\epsilon}}
-\frac{i}{\sqrt{-x^2+i\epsilon}}\frac{i}{\pi}\frac{2}{(-im\sqrt{-x^2+i\epsilon})}\nonumber\\
&=&\frac{2i}{m\pi(x^2-i\epsilon)}+\frac{2i}{m\pi(-x^2+i\epsilon)}\nonumber\\
&=&0
\end{eqnarray}
The reason why the authors of Ref.\cite{Greiner:1992bv} regarded the
above term as to be nonzero may come from that they had taken both
$x^2$ and $-x^2$ to be the absolute value $|x^2|$ simultaneously in
their calculation. However, it is obviously impossible that both
$x^2$ and $-x^2$ were equal to $|x^2|$, even if $x^2\to 0$, because
$x^2$ can only be approaching zero from either of the positive or
the negative axis direction, that is, in any case $x^2$ and $-x^2$
always have opposite signs even if they are infinitesimal.

Moreover, from eq.\eqref{eq-Dp1-Dirac} together with
eq.\eqref{eq-Dp1-summary}, we obtain an interesting recurrence
relation for the spinor Feynman propagator in $(D+1)$-dimensional
spacetime and the scalar Feynman propagators in $(D-1)$-, $(D+1)$-
and $(D+3)$-dimensional spacetimes as follows
\begin{eqnarray}
S_F^{(D+1)}(x)=\bigg(-\frac{i\slashed{x}}{x^2-i\epsilon}+m\bigg)D_F^{(D+1)}(x)
-\frac{im^2\slashed{x}}{4\pi(x^2-i\epsilon)}D_F^{(D-1)}(x)
+i\pi\slashed{x}D_F^{(D+3)}(x)
\end{eqnarray}
where the superscripts denote the spacetime dimensions of the
respective physical quantities. The above relation essentially stems
from the facts that the Feynman propagator in any dimensional
spacetime can be expressed in terms of Bessel and modified Bessel
functions, which has been proved in this paper. And it exhibits that
the free Dirac theory and the free Klein-Gordon theories in
alternate-successive dimensional spacetimes might be related to each
other.

\section{Conclusions\label{section-conclusion}}

In this paper, we have pointed out and corrected an error of the
results of previous works on the analytic expression of the Feynman
propagator in position space of a Dirac spinor in
$(3+1)$-dimensional spacetime, and we have derived the generalized
analytic formulas of both the scalar Feynman propagator and the
spinor Feynman propagator in position space in any
$(D+1)$-dimensional spacetime. The method we have used in this paper
is different from that used in Ref.\cite{Greiner:1992bv}. And the
result we have obtained shows that the analytic formula of the
Feynman propagator in position space can be also expressed in terms
of Hankel functions of second kind and Modified Bessel functions in
a general $(D+1)$-dimensional spacetime, just like the known case in
$(3+1)$-dimensional spacetime. From the obtained results, at the end
we have found an interesting recurrence relation among the spinor
Feynman propagator in $(D+1)$-dimensional spacetime and the scalar
Feynman propagators in $(D+1)$-, $(D-1)$- and $(D+3)$-dimensional
spacetimes.

\begin{acknowledgments}
We would like to thank Profs. Qing Wang, Zhan Xu, Qiong-Gui Lin, and
Nai-Ben Huang for helpful discussions. This work is supported by the
National Natural Science Foundation of China, and Sun Yet-Sen
University Science Foundation.
\end{acknowledgments}

\end{document}